\documentclass{egs}
\usepackage{epsfig}

\printfigures              

\begin{document}
\title{Model Bond albedos of extrasolar giant planets}

\author{Christopher R. Gelino}
\author{M. Marley}
\author{D. Stephens}
\affil{Department of Astronomy, New Mexico State University, Las Cruces, New Mexico}
\author{J. Lunine}
\affil{Lunar and Planetary Laboratory, University of Arizona, Tucson, Arizona and Reparto di Planetologia, Istituto di Astrophysica, Rome, Italy}
\author{R. Freedman}
\affil{Space Physics Research Institute, NASA/Ames Research Center, Moffett Field, California}

\date{December 8, 1998}

\journal{\PCE}       

\firstauthor{Gelino}
\proofs{Christopher R. Gelino\\Department of Astronomy\\MSC 4500\\New Mexico State University\\P.O. Box 30001\\Las Cruces, NM 88003-8001}
\offsets{Christopher R. Gelino\\Department of Astronomy\\MSC 4500\\New Mexico State University\\P.O. Box 30001\\Las Cruces, NM 88003-8001\\email: crom@nmsu.edu}

\msnumber{PS12-016}
\received{}
\accepted{}

\maketitle

\begin{abstract}
The atmospheres of extrasolar giant planets are modeled with various effective temperatures and gravities, with and without clouds.  Bond albedos are computed by calculating the ratio of the flux reflected by a planet (integrated over wavelength) to the total stellar flux incident on the planet.  This quantity is useful for estimating the effective temperature and evolution of a planet.  We find it is sensitive to the stellar type of the primary.  For a 5 $M_{\rm Jup}$ planet the Bond albedo varies from 0.4 to 0.3 to 0.06 as the primary star varies from A5V to G2V to M2V in spectral type.  It is relatively insensitive to the effective temperature and gravity for cloud--free planets.  Water clouds increase the reflectivity of the planet in the red, which increases the Bond albedo.  The Bond albedo increases by an order of magnitude for a 13 $M_{\rm Jup}$ planet with an M2V primary when water clouds are present.  Silicate clouds, on the other hand, can either increase or decrease the Bond albedo, depending on whether there are many small grains (the former) or few large grains (the latter). 
\end{abstract}

\section{Introduction}\label{sec:intro}
Within the past 4 years there has been a wealth of indirect
evidence that planets exist around other stars.  These include planetary companions to 51 Pegasi \citep{mq95}, 47 Ursae Majoris \citep{bm96}, and 70 Virginis \citep{mb96} to name a few.  To date, there are about a dozen confirmed planets around main sequence stars.  These planets cover a wide range of masses ($\sim$ 0.5 $M_{\rm Jup}$ -- $\sim$ 10 $M_{\rm Jup}$, where 1 $M_{\rm Jup}$ = 1 Jupiter mass = 1.9 $\times$ 10$^{27}$ kg) and orbital semi-major axes ($\sim$ 0.04 AU -- 2.5 AU).  The variety of semi--major axes (all of which are smaller than Jupiter's semi--major axis of 5.2 AU) and ages of planets' primary stars means that these planets also have a large spread of temperatures.

The reflectivity of a planet is important for understanding the nature and the evolution of that planet.  This quantity depends on the abundance of absorbers and scatterers that are present in the atmosphere, which in turn depend on the pressure--temperature profile and thermal history of the planet.  From the example of planets in our own solar system, we know that clouds play an important role in the reflectivity. 

This paper is a companion to \citet{marl99}.  Whereas Marley et al. focus on the reflected spectra and the effects of different cloud particle and supersaturation assumptions, here we summarize the atmospheric models of Marley et al. and \citet{bur97} and focus on the sensitivity of the Bond albedo with various parameters, such as the effective temperature and gravity of the extrasolar giant planet or brown dwarf (hereafter, the abreviation EGP is used to describe these objects, regardless of their origin or mass), and the spectral type of the primary star.

The models of \citet{bur97} are used for the pressure­-temperature profiles of EGPs with effective temperatures between 100 and 1200 K and gravities between 5 and 3000 m/s$^2$.  Models assume elemental solar abundances and thermochemical equilibrium is used to determine the abundances of the primary opacity sources in the atmosphere: H$_2$O, NH$_3$, CH$_4$, CO, H$_2$--He, and H$_2$--H$_2$.  The scattering and extinction properties are calculated for these sources as a function of wavelength.  See \citet{marl99} for the complete details of the models.

Since the planets vary in distance from their primary star, the geometric albedo is used to characterize their reflectivities.  The geometric albedo, $p$, is the percent of the incident light reflected at zero phase angle.  It does not depend on either the wavelength dependence or the intensity of the incident light and is, therefore, very useful.

The Bond albedo, $A$, describes the total reflectivity of the planet.  It takes into account the geometric albedo, the phase integral, and the light incident upon the planet.  It is the ratio of the total reflected power, $P_{refl}$, to the total incident power, $P_{incid}$:

\begin{equation}
 A = \frac{P_{refl}}{P_{incid}} = \frac{\int^\infty_0 F(\lambda) q(\lambda) p(\lambda) d\lambda}{\int^\infty_0 F( \lambda ) d\lambda} \label{eq:A},
\end{equation}

\noindent
where $F(\lambda)$ is the incident stellar flux and $q(\lambda)$ is the phase integral, both of which are functions of wavelength $\lambda$.  

\section{Opacity Sources}\label{sec:opsrc}
Full treatment of the opacity sources is given in \citet{marl99}.  Here we summarize the important sources and processes.

The dominant opacity sources in the infrared (IR) for these atmospheres are H$_2$O, CH$_4$, NH$_3$, CO, and pressure--induced molecular hydrogen opacity.  Their relative abundances are determined by assuming solar abundances of the elements and by computing thermochemical equilibrium.

Scattering processes also contribute to the opacity.  The cross section of Rayleigh scattering by atoms and molecules varies as 1/$\lambda^4$.  Thus, this process is very efficient at short $\lambda$.  Longer wavelength photons, however, are absorbed in the atmosphere before they are Rayleigh scattered.  As a result, Rayleigh scattering generally occurs high in the atmosphere.  In addition to this process, Raman scattering, in which short wavelength photons are scattered to longer wavelengths, is also included.  We assume that Raman scattered photons are removed from the system.

Mie scattering is the scattering of radiation off spherical particles and grains with sizes comparable to the wavelength of the incident photons.  This theory is used to compute the scattering properties of the clouds described in Sect.~\ref{sec:clouds}.

\section{Clear atmosphere models}\label{sec:clear}
The pressure­-temperature profiles of \citet{bur97} were computed for EGPs with a given effective temperature and gravity and a clear atmosphere.  These models are  self­-consistent for isolated EGPs.  However, to derive a Bond albedo and a reflected spectrum, light from a primary star must be reflected off the object.  This stellar radiation would heat the planet and change the pressure--temperature profile.  The amount that the profile would change depends on the stellar type of the primary and the distance between the planet and the star.  These models do not currently include heating effects from the primary star.  Recently, \citet{seag98} have shown that the emergent spectra of EGPs does change compared to that for an isolated EGP.  For a dust--free model with $T_{\rm eff}$ = 1835 K, the differences are typically less than an order of magnitude between 0.5 and 5 $\mu$m.  For the cooler models considered here, the influence of the incident sunlight should be less than that found by Seager and Sasselov for the much hotter models.

Based on the opacity sources in Sect.~\ref{sec:opsrc}, the geometric albedos and phase integrals are calculated as a function of wavelength for each EGP.  Reflected spectra and Bond albedos are then computed from the geometric albedo, the phase integral, and the spectrum of the primary star.

\section{Cloudy atmosphere models}\label{sec:clouds}
The formalism of \citet{ros78} is used to compute cloud structure and particle size distributions.  The theory starts with an initial distribution of  ``molecular embryos,'' which are allowed to grow by nucleation (the formation of droplets from the embryos), condensation (the growth of droplets through the inward flux of vapor molecules), and coagulation (the growth of droplets through collisions). In addition, droplets are removed from the cloud by evaporation (the outward flux of molecules through a droplet's surface) and sedimentation (the falling of droplets under the influence of gravity).  We further consider two situations for the vertical distribution of the clouds.  For a more detailed discussion of the cloud models see \citet{marl99}.

A quiescent atmosphere is one in which turbulence does not generate macroscopic eddy motions which serve to keep relatively large particles from rapidly sedimenting out of the cloud layer.  In this atmosphere we balance, as a function of particle size, the growth rate against the sedimentation rate.  Particles large enough such that the sedimentation rate just exceeds the growth rate are assumed to be the modal particle size.  

For a turbulent atmosphere the same procedure is followed, but here the rate of eddy mixing is computed and balanced against the sedimentation rate as a function of particle size.  The turbulent motions of eddy mixing provide an upward lift that opposes gravity, consequently lowering the sedimentation rate.  The turbulent motions also mix the particles in the cloud, allowing them to collide and create larger particles.  As a particle grows in radius and mass, the sedimentation rate increases and the eddy mixing rate decreases.  Particles large enough that the sedimentation rate just exceeds the rate of remixing by eddy turbulence are again assumed to fall out of the cloud.  The resulting particle sizes as a function of altitude are significantly larger than for the quiescent cloud.   

Two cloud species are incorporated, water and enstatite (MgSiO$_3$), whose existence depends on the ambient atmospheric temperature.  For a 2 $M_{\rm Jup}$ EGP with $T_{\rm eff}$ = 300 K, water starts to condense at an atmospheric temperature around 225 K and pressure around 0.07 bars \citep{bur97}.  Water condenses lower in the atmosphere as $T_{\rm eff}$ increases.  For EGPs with $T_{\rm eff}$ $>$ 400 K the atmospheric temperatures are too high for water clouds to condense.

Enstatite starts to condense at much higher temperatures ($T$ $\sim$ 1800 K) and pressures ($\approx$ 48 bars) for 36 $M_{\rm Jup}$ mass EGPs at $T_{\rm eff}$ = 1000 K \citep{bur97}.  The pressures at which these clouds form for EGPs with $T_{\rm eff}$ $<$ 1000 K is very deep ( $>$ 300 bars).  The conditions for enstatite condensation in lower mass EGPs occurs very deep in the atmosphere.  Therefore, these clouds are only important for EGPs at $T_{\rm eff}$ $>$ 1000 K.   

We consider one silicate and two water cloud models which depend on the EGP's effective temperature and gravity.  The silicate cloud is calculated for a 36 $M_{\rm Jup}$ planet at $T_{\rm eff}$ = 1000 K; the water clouds are calculated for 2 and 13 $M_{\rm Jup}$ EGPs at $T_{\rm eff}$ = 300 K.  Each of these models is computed in both a quiescent and a turbulent atmosphere, giving a total of six cloud models (two silicate, four water).

\section{Results}\label{sec:res}
\begin{figure}
\centerline{\epsfig{file=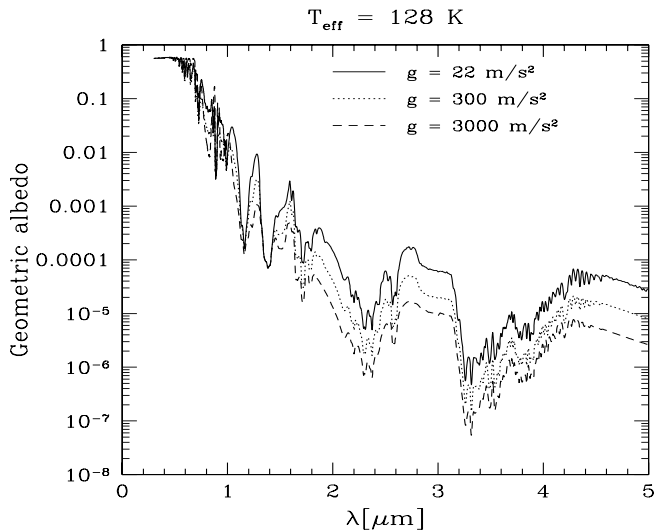, width=9cm, height=7.5cm}}
  \caption[]{\label{fig:geo}
The geometric albedo for clear atmosphere EGPs with $T_{\rm eff}$ = 128 K, and g = 22 (solid), 300 (dotted), and 3000 (dashed) m/s$^2$.  These correspond to objects with masses of roughly 2, 10, and 45 $M_{\rm Jup}$.}
 \end{figure}
 
Even a cursury look at an EGP's geometric albedo (Fig. \ref{fig:geo}) indicates that EGPs are not grey reflectors (reflect the same percentage of radiation at all $\lambda$).  As mentioned above, Rayleigh scattering is most efficient at small $\lambda$.  This effect is clearly seen in the geometric albedo.  All clear atmosphere EGPs show an increase in $p$ towards small $\lambda$.  Absorption by molecules in the IR causes a dramatic decrease (by about 5 orders of magnitude) from 0.5 $\mu$m to 3.2 $\mu$m.  This result has important consequences when the spectra of stars of different spectral types (and hence temperatures) are reflected off these EGPs (see below).

\begin{figure}
\centerline{\epsfig{file=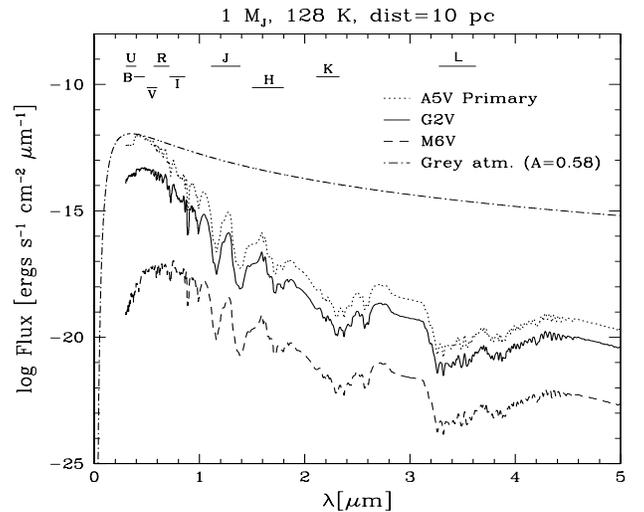, width=9cm, height=7.5cm}}
  \caption{\label{fig:spec}
Reflected spectrum of a cloud-free EGP with M = 1 $M_{\rm Jup}$ and $T_{\rm eff}$ = 128 K, placed 1 AU from primary stars of spectral type A5V (dotted), G2V (solid), and M6V (dashed).  The dot-dashed curve is the reflected flux for a planet that is a grey reflector with an albedo of 0.58.  Several photometric filter bandpasses are shown for reference.}
\end{figure}

Figure~\ref{fig:spec} shows how the light reflected off a 1 $M_{\rm Jup}$ planet (for this and all subsequent figures the EGP--to--star distance is 1 AU) changes with the spectral type of the primary.  For the most part the shapes of the spectra are similar---the small differences at short $\lambda$ are real features in the stellar spectra.  Also, these spectra have not been shifted for clarity.  Thus, the magnitudes of the fluxes are also properties of the different spectral types.  The hottest star, A5V, has a higher intrinsic flux and shorter peak wavelength than the cooler stars, G2V and M6V.  The dot-dashed curve is the resulting spectrum if the planet is a grey reflector with an albedo of 0.58 and is reflecting the light from a perfect blackbody with $T_{\rm eff}$ = 8300 K ($T_{\rm eff}$ for an A5V star).  Since the spectra of the stars resembles that of a blackbody, many differences seen in the spectra of the grey reflector and the EGP spectra are caused by the planets themselves.  Clearly, EGPs are much darker in the IR than what is expected if they were grey reflectors.

\begin{figure}
\centerline{\epsfig{file=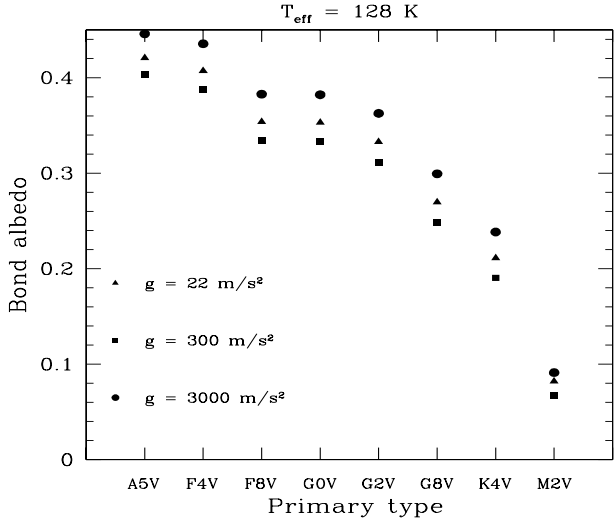, width=9cm, height=7.5cm}}
  \caption[]{\label{fig:bond}
The Bond albedo as a function of primary spectral type for clear atmosphere EGPs with $T_{\rm eff}$ = 128 K, and g = 22 (triangles), 300 (squares), and 3000 (circles) m/s$^2$.}
\end{figure}

Not only is the shape of the geometric albedo for an EGP important for the calculation of $A$, but also the shape of the stellar spectrum.  In general the Bond albedo of an EGP with a hotter primary (A and F) is greater than the Bond albedo for planets with cooler primaries (M--type, Fig.~\ref{fig:bond}).  The hotter stars peak in emitted flux in the blue, where EGPs are the most reflective.  The cooler stars, on the other hand, peak at longer $\lambda$, where EGPs are darker.  Because of the dependence of $A$ on $p(\lambda)$ and $F(\lambda)$ (Eq.~\ref{eq:A}), these combinations lead to bright EGPs around hot stars and dark EGPs around cool stars.  In addition, the Bond albedo does not depend on the brightness of the star, only the shape of the spectrum.

One important caveat is that these atmosphere models do not contain photochemical-produced species (e.g. C$_2$H$_2$) that absorb strongly in the UV.  The presence of such species will lower the reflectivity in the blue and UV and consequently lower the Bond albedos for the planets with bluer primaries.

\begin{figure}
\centerline{\epsfig{file=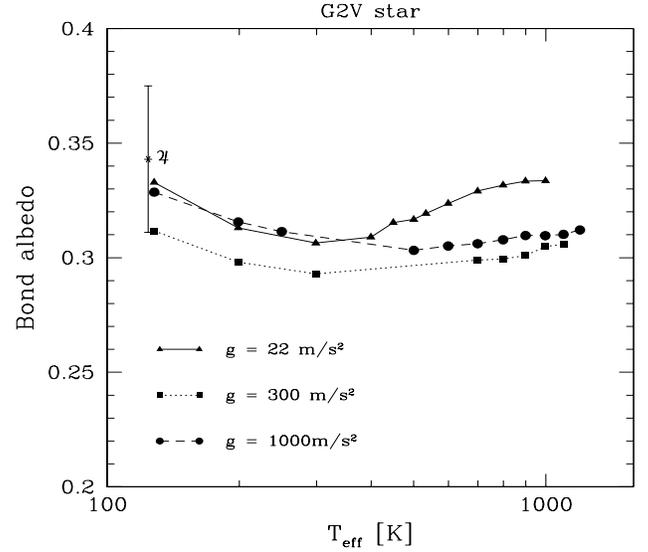, width=9cm, height=8cm}}
  \caption{\label{fig:temp}
The Bond albedo as a function of $T_{\rm eff}$ for clear atmosphere planets with g = 22 (triangles), 300 (squares), and 1000 (circles) m/s$^2$.  The Bond albedo of Jupiter is plotted for comparison.}
\end{figure}

\begin{figure}
\centerline{\epsfig{file=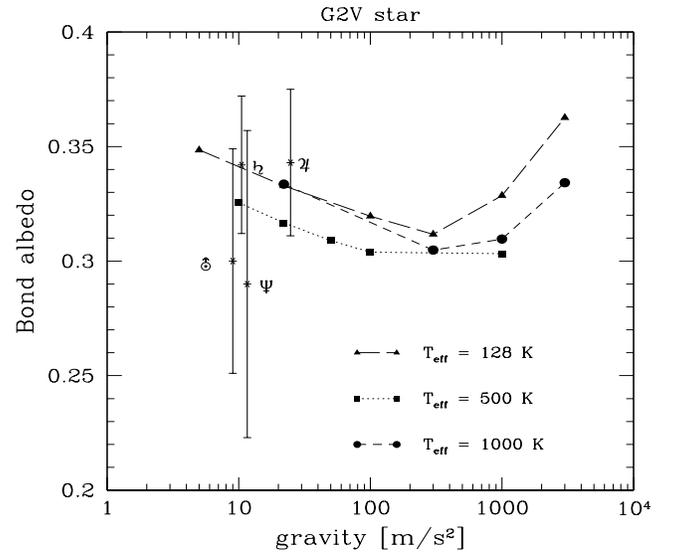, width=9cm, height=8cm}}
  \caption{\label{fig:grav}
The dependence of the Bond albedo on the gravity of clear atmosphere planets with $T_{\rm eff}$ = 128 (triangles), 500 (squares), and 1000 (circles) K.  The Bond albedos of Jupiter, Saturn, Uranus, and Neptune (counter--clockwise from upper right; Conrath et al., 1989) are plotted for comparison.}
\end{figure}

The Bond albedo as a function of $T_{\rm eff}$ of EGPs is shown in Fig.~\ref{fig:temp}.  Despite approximately 2 orders of magnitude difference in their gravities, EGPs with a $T_{\rm eff}$ close to that of Jupiter have Bond albedos that are within the error bars of Jupiter's albedo.  The Bond albedo is, therefore, relatively insensitive to the gravity of clear atmosphere EGPs.  $A$ is also insensitive to $T_{\rm eff}$ for EGPs (Fig.~\ref{fig:grav}).  The Bond albedos of the solar system giants are comparable to those of EGPs with gravities similar to those of the giant planets.  

The reflectance properties of the cloud--free EGPs discussed above are only limiting cases.  Indeed, given the complexity of clouds seen in the planets of our solar system, we expect a rich variety of cloud properties in the new objects.  Therefore, we consider the effects that clouds have on the reflectivity of EGPs.

\begin{figure}
\centerline{\epsfig{file=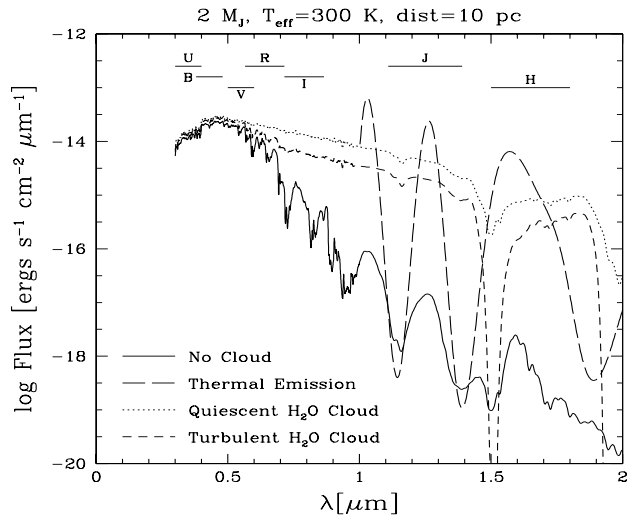, width=9cm, height=7.5cm}}
  \caption{\label{fig:wspec}
The reflected spectrum for a planet with M = 2 $M_{\rm Jup}$, $T_{\rm eff}$ = 300 K, with a G2V primary, which is 10 pc from the Earth.  The turbulent and quiescent H$_2$O cloud models (short dashed and dotted curves respectively) can enhance the reflected flux by several orders of magnitude over the clear atmosphere model (solid curve).}
\end{figure}

Water clouds can play a large role in the reflectivity of EGPs with $T_{\rm eff}$ $\leq$ 400 K.  In particular, the reflectivity for a 2 $M_{\rm Jup}$ ($T_{\rm eff}$ = 300 K) EGP is increased in the IR, as seen in Fig.~\ref{fig:wspec}.  At short $\lambda$ the incident photons are Rayleigh scattered high in the atmosphere, before the photons reach the cloud layer.  The water clouds are high enough in the atmosphere (pressure $<$ 0.1  bars) that the longer wavelength (IR) photons which do reach them are not severely attenuated.  The result is that water clouds brighten this EGP in the IR by several orders of magnitude.

In addition to the reflected flux, an EGP also emits thermal radiation.  The thermal emission shown in Fig.~\ref{fig:wspec} is for a clear atmosphere EGP with $T_{\rm eff}$ = 300 K.  Like the geometric albedo, the thermal emission is non--grey \citep{bur97}.  When considering cloud--free planets, the thermal emission for this $T_{\rm eff}$ dominates over the reflected flux.  The formation of water clouds provides enough of a boost in the reflectivity that the reflected flux is raised to intensities comparable to the thermal flux.

\begin{figure}
\centerline{\epsfig{file=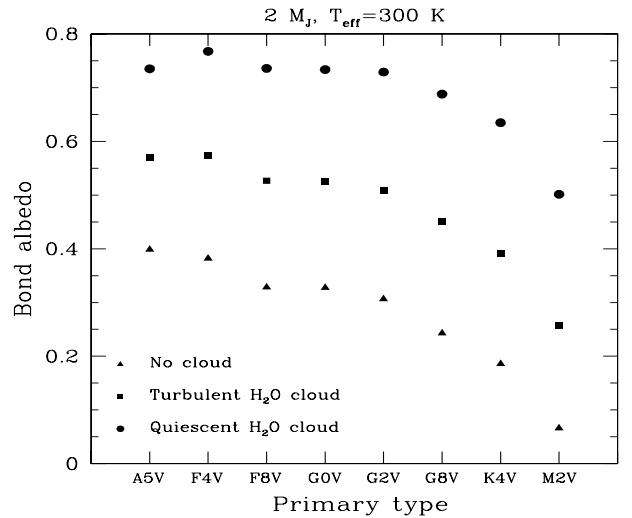, width=9cm, height=7.5cm}}
  \caption{\label{fig:wbonds}
The Bond albedo as a function of primary stellar type for planets with M = 2 $M_{\rm Jup}$, $T_{\rm eff}$ = 300 K, and which include no clouds (triangles), turbulent H$_2$O clouds (squares), and quiescent H$_2$O clouds (circles).}
\end{figure}

As seen in Fig.~\ref{fig:wbonds}, H$_2$O clouds can increase the Bond albedo of this model by as much as an order of magnitude in some cases (M2V spectral type primary).  In general, the Bond albedos for planets around cool stars are enhanced more than for planets around hotter stars when water clouds are added.  Again, the cooler stars peak in their emitted flux at longer $\lambda$, where the planets with clouds have relatively high reflectivities.

The vastly different Bond albedos for the turbulent and quiescent clouds are due to the number of scatterers in each cloud type.  The amount of water at a given pressure level in the atmosphere is the same for quiescent clouds as it is for the turbulent clouds, but quiescent cloud particles are smaller than turbulent cloud particles.  Thus, for a given level in the atmosphere, there are many more scatterers present in the quiescent cloud than in the turbulent cloud.  The result is higher geometric albedos and Bond albedos for the quiescent clouds.  In addition, these calculations also depend on the cloud supersaturation, which is discussed in \citet{marl99}.  Therefore, the quiescent cloud probably gives an upper limit to the actual Bond and geometric albedos.

\begin{figure}
\centerline{\epsfig{file=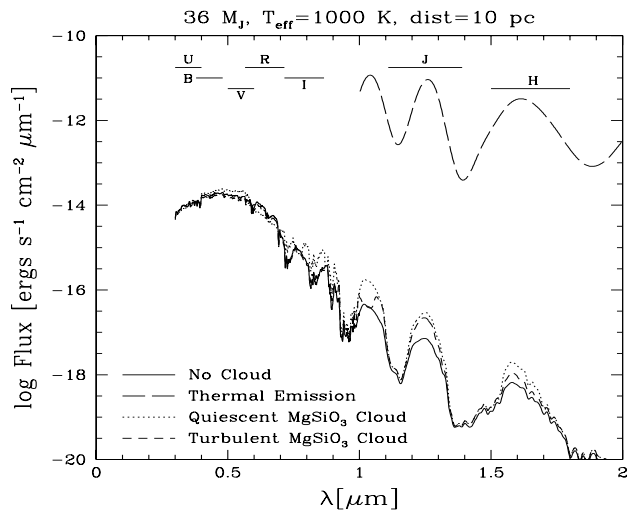, width=9cm, height=7.4cm}}
  \caption{\label{fig:sspec}
The reflected spectrum from an EGP with M = 36 M$_{Jup}$, $T_{\rm eff}$ = 1000 K, and positioned 1 AU from a G2V star, which is 10 pc from the Earth.  The MgSiO$_3$ clouds have only a small effect over cloud-free models (solid) in the reflected flux.}
\end{figure}

Enstatite (MgSiO$_3$) clouds in hotter EGPs have a much smaller effect on the reflected flux than do the water clouds (Fig.~\ref{fig:sspec}).  The spectrum is relatively unchanged when either quiescent MgSiO$_3$ clouds or turbulent MgSiO$_3$ clouds are present in this model.  These clouds form at pressures $\sim$48 bars, which is considerably deeper than the water clouds.  As with the water clouds, most of the blue photons from the star are scattered high in the atmosphere via Rayleigh scattering.  Most of the longer wavelength photons are absorbed or scattered in the atmosphere before penetrating to the depth of the cloud layer.  There are a few IR regions (around 1, 1.25 and 1.6 $\mu$m), however, where the gas optical depth is small enough that photons are reflected off the clouds.  In these regions the reflected flux is enhanced by an order of magnitude at most.  

\begin{figure}
\centerline{\epsfig{file=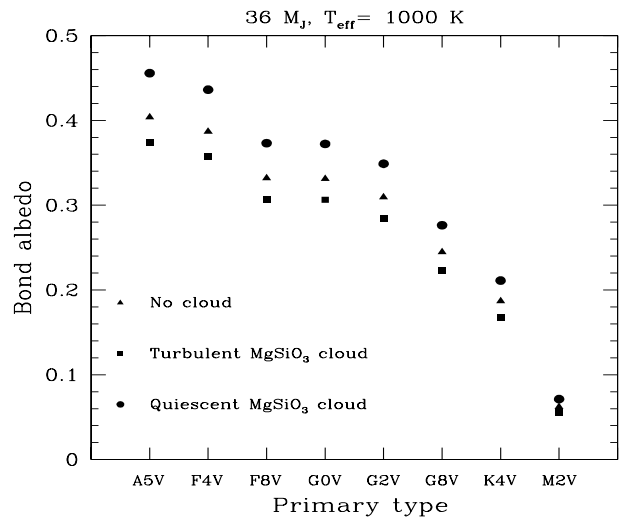, width=9cm, height=7.4cm}}
  \caption{\label{fig:sbonds}
Same as for Fig.~\ref{fig:wbonds} except for M = 36 $M_{\rm Jup}$, $T_{\rm eff}$ = 1000 K, and MgSiO$_3$ clouds are present.}
\end{figure}

The Bond albedos for these EGPs as a function of spectral type are shown in Fig.~\ref{fig:sbonds}.  Like the quiescent H$_2$O clouds, the quiescent MgSiO$_3$ clouds also produce the largest Bond albedos.  The turbulent MgSiO$_3$ clouds, on the other hand, produce albedos that are smaller than those for the clear atmosphere case.  This difference is attributable to the differing column abundances and scattering properties of the two cloud models.  The phase integral is considerably lower for the turbulent cloud than it is for the quiescent cloud.  This, in turn, lowers the Bond albedo.

\section{Conclusions}
We have modeled the atmospheres of EGPs with effective temperatures between 100  and 1200 K and gravities between 5 and 3000 m/s$^2$.  We have assumed solar abundances of the elements and computed thermochemical equilibrium to obtain the abundances of the primary molecular opacity sources.  The effects of heating by the primary star and reflection of stellar radiation from the clouds are neglected when calculating the pressure--temperature profiles of the planets.  Reflected spectra and Bond albedos are computed for EGPs with and without clouds.


The conclusions can be summarized as follows:

\noindent
(1) The reflected spectra of EGPs are not grey.  Rayleigh scattering makes them very reflective in the blue part of the spectrum and molecular opacities make them darker (by about 5 orders of magnitude) in the IR than the visible for clear atmospheres.

\noindent
(2) The Bond albedo is a strong function of the spectral type of the primary star.  A Jupiter mass planet orbitting a hot A5V star reflects $\sim$4 times as much of the incident radiation as that planet does around a cooler M2V primary.

\noindent
(3) The Bond albedos of clear atmosphere EGPs are relatively insensitive to the body's gravity and effective temperature.

\noindent
(4) Thermal emission dominates the IR spectrum for $T_{\rm eff}$ $>$ 400 K.  At smaller $T_{\rm eff}$, H$_2$O clouds can be present and can boost the reflected spectrum to levels comparable to the thermal emission (see below).

\noindent
(5) For EGPs with $T_{\rm eff}$ $<$ 400 K water clouds can dramatically brighten the reflectivity over that for a cloud--free EGP.  Hotter planets do not have water clouds and, therefore, would not show bright IR reflected fluxes.

\noindent
(6) Condensation of silicate grains occurs too deeply in the atmosphere for objects with $T_{\rm eff}$ $<$ 1000 K to substantially affect the reflected flux
or the Bond albedo.  For larger effective temperatures the grains condense higher in the atmosphere and will play a more important role.  \citet{seag98}  discuss the role of grains for very hot ($\sim$1800K) objects.


\begin{acknowledgements}
The authors wish to thank the anonymous referees for their useful comments.
\end{acknowledgements}

\end{document}